\input harvmac
\noblackbox

\def\ts{\theta_6}

\def\tiny{\scriptscriptstyle\rm}

\def\free{\hbox{$\cal F$}}
\def\bold#1{\setbox0=\hbox{$#1$}%
     \kern-.010em\copy0\kern-\wd0
     \kern.025em\copy0\kern-\wd0
     \kern-.020em\raise.0200em\box0 }
\def\nabb{\bold{\nabla}}

\def\dz{\partial_z}

\def\di{\partial_i}

\def\cross{\!\times\!}

\lref\DG{P.G.~de~Gennes, {Solid State Commun.} {\bf 14} (1973) 997.}

\lref\KasandMac{F.C.~MacKintosh, J.~Kas and P.A.~Janmey, Phys. Rev. Lett. {\bf 75}
(1995) 4425.}

\lref\Fradenii{Z.~Dogic and S.~Fraden, Phys. Rev. Lett.
{\bf 78} (1997) 2417.}

\lref\Safin{J.O.~R\"adler, I.~Koltover, T.~Salditt and C.R.~Safinya,
Science {\bf 275} (1997) 810.}

\lref\Tirrell{G.H.~Zhang, M.J.~Fournier, T.L.~Mason and 
D.A.~Tirrell, Macromolecules {\bf 25} (1992) 3601.}

\lref\Felgner{P.L.~Felgner and G.~Rhodes,
Nature {\bf 349} (1991) 351.}

\lref\Safinya{H.E.~Warriner, S.H.J.~Idziak, N.L.~Slack, P.~Davidson and
C.R.~Safinya, Science {\bf 271} (1996) 969;  S.L.~Keller, H.E.~Warriner, C.R.~Safinya
and J.A.~Zasadzinski, Phys. Rev. Lett. {\bf 78} (1997) 4781.}

\lref\TER{E.M.~Terentjev, {Europhys. Lett.} {\bf 23} (1993 27).}

\lref\TON{J.~Toner, {Phys. Rev. A} {\bf 27} 1157 (1983) 1157.}

\lref\GIA{C.~Gianessi, {Phys. Rev. A} {\bf 28} (1983) 350; 
{\bf Phys. Rev. A} {\bf 34} (1986) 705.}

\lref\LIVO{
F.~Livolant, {Physica A} {\bf 176} (1981) 117.}

\lref\HN{B.I.~Halperin and D.R.~Nelson, {Phys. Rev. Lett.}
{\bf 41} (1978) 121;
D.R.~Nelson
and B.I.~Halperin, {Phys. Rev. B} {\bf 19} (1979) 2457.}

\lref\PN{P.~Nelson and T.~Powers, {Phys. Rev. Lett.}
{\bf 69} (1992) 3409;
{\bf J. Phys. II (Paris)} {\bf 3} (1993) 1535.}

\lref\TGB{S.R.~Renn and T.C.~Lubensky, {Phys. Rev. A}
{\bf 38} (1988) 2132; T.C.~Lubensky
and S.R.~Renn, {Phys. Rev. A} 
{\bf 41} (1990) 4392.}

\lref\MEYER{R.B.~Meyer, {Appl. Phys. Lett.},
{\bf 12}, 281 (1968); {\bf 14}, 208
(1969).}

\lref\FEL{L.G.~Fel, {Phys. Rev. E}, {52}, 702 (1995).}

\lref\HKL{A.B.~Harris, R.D.~Kamien and T.C.~Lubensky, Phys. Rev. Lett. 1476 
{\bf 78}
(1997); {\bf 78} (1997) 2867.}

\lref\LCBO{R.D.~Kamien, {J. Phys. II France} {\bf 6} 
(1996) 461.}

\lref\KN{R.D.~Kamien and D.R.~Nelson, {Phys. Rev. Lett.}
{\bf 74} (1995) 2499;
{Phys. Rev.} {\bf E} {\bf 53} (1996) 650.}

\lref\STREY{R.~Podgornik, H.H.~Strey, K.~Gawrisch, D.C.~Rau,
A.~Rupprecht and V.A.~Parsegian,  Proc.
Nat. Acad. Sci. {\bf 93} (1996) 4261.}

\lref\KT{R.D.~Kamien and J.~Toner, Phys. Rev. Lett. {\bf 74} (1995) 3181.}

\lref\BOUii{F.~Livolant
and Y.~Bouligand, J. Phys. (Paris) {\bf 47} (1986) 1813; A.~Le~Forestier and
F.~Livolant, Liq. Cryst. {\bf 17} (1994) 651; see also
D.C.~Martin and E.L.~Thomas, Phil. Mag. A {\bf 64} (1991) 903.}

\lref\IND{V.L.~Indenbom and A.N.~Orlov, Usp. Fiz. Nauk {\bf 76} (1962) 557 [
Sov. Phys. Uspekhi {\bf 5} (1962) 272].}
\lref\KOS{A.M.~Kosevich, Usp. Fiz. Nauk {\bf 84} (1964) 579 [Sov. Phys.
Uspekhi
{\bf 7} (1965) 837].}

\lref\TLL{R.D.~Kamien and T.C.~Lubensky, J. Phys. I France {\bf 3}
(1993) 2131.}

\lref\FLUX{R.D.~Kamien, Phys. Rev. B {\bf 58} (1998) 8218.}

\lref\ILCC{R.D.~Kamien, Mol. Cryst. Liq. Cryst. {\bf 299} (1997) 265.}

\nfig\fone{The moir\'e state.  The thick tubes
running in the $\hat z$ direction are polymers, while the dark lines are
stacked honeycomb arrays of screw dislocations.  The intersection of
these polymers with any constant $z$ cross section away from the hexagonal
defect arrays has the topology of a triangular lattice.}

\nfig\ftwo{Phase diagram of a chiral polymer crystal.  Insets are
representative
tilt (TGB) and moir\'e grain boundaries.  Shaded lines are screw dislocations.
In the TGB phase the solid lines are the polymers in front of the 
grain-boundary, while the dashed ones are behind it.  In the moir\'e phase
the crosses are the heads of polymers beneath the grain-boundary and 
the circles are the tails of the polymers above it.}

\nfig\fthree{X-ray structure function in the plane perpendicular
to the nematic direction \STREY.  This diffraction pattern contains
a non-zero $\cos 6\theta$ component and no measurable $\cos 6n\theta$ for $n\ge 2$.
The small amount of $\cos 2\theta$ can be attributed to the misalignment
of the X-ray beam.  (Figure provided courtesy of R.~Podgornik).}

\nfig\fff{Model of a chiral hexatic.  In each plane the bond-order
parameter is $\ts=\ts^0\;{\rm mod}\;2\pi/6$.  Between the planes the bond
order uniformly precesses along the average nematic director, ${\bf\hat n}
=\hat z$.  The planes are analogous to smectic planes, though there is no
density wave in this liquid crystalline phase.}

\Title{}{Chiral Mesophases of DNA}

\centerline{
Randall D. Kamien\footnote{$^\dagger$}{email: {\tt
kamien@physics.upenn.edu}} }

\smallskip\centerline{\sl Department of Physics and
Astronomy, University of Pennsylvania, Philadelphia,
PA 19104}

\vskip .3in
In the hexagonal columnar phase of chiral polymers a bias towards cholesteric
twist competes with braiding along an average direction.  When the chirality
is strong, topological defects proliferate, leading to either a tilt grain
boundary phase or a new ``moir\'e state'' with twisted bond order.
This moir\'e phase can melt leading to a new phase: the chiral hexatic.
I will discuss some recent experimental results from the NIH on DNA liquid
crystals in the context of these theories.

\vfill{\bf To appear in the Proceedings of 
the $35^{\rm
th}$ Annual Technical Meeting of the Society of Engineering Science, Pullman, WA, 
27-30 September 1998.}
\Date{8 December 1998}
\newsec{Introduction and Summary}
Chiral molecules are ubiquitous in nature \LIVO .  It is remarkable, in
fact, that
cellular processes can produce copious amounts of chiral molecules of
the same handedness, while, by comparison, standard synthetic techniques
usually produce racemic mixtures in which there are an equal
number of left- and right-handed molecules.  While each molecule in a racemic
mixture
can be chiral, from the point of view of coarse-grained, effective models
these systems are not chiral.  Thus studying biologically produced
materials is intriguing:  it allows one to consider macroscopic chiral
structures composed of regular, sometimes monodisperse, molecules.
It is already known that these materials have remarkable elastic properties --
for instance, they become {\sl more} viscous on dilution \Safinya\ and have
remarkable viscoelastic properties because of their rigidity \KasandMac .
Moreover, to a theorist these biomaterials offer a significantly cleaner
environment since they can be produced with greater regularity and
monodispersity \refs{
\Tirrell,\Fradenii}
than is possible for conventional polymers.
Thus theoretical discussions and
predictions
of the
effects of polydispersity, chirality and interaction types ({\sl i.e.} steric,
van der Waals,
screened Coulomb, {\sl etc.}) can be made, for the first time,
in the context of experiment.   Because biomolecules are the building blocks of
ultrastrong materials, such as silkworm and spider silk, theoretical
understanding
of the allowed and preferred structures is an essential element in the design
of strong
materials based on molecular constituents.
In addition, DNA-lipid complexes \Safin\ show promise as therapeutic
DNA transfection systems that do not require virus vectors \Felgner .
Again, the detailed structure and packing of the biomolecules in these
complexes is
crucial for advances in successful DNA delivery.

In the following, I shall describe three new chiral liquid crystalline
phases, which one might expect to observe in these relatively pure,
biomolecular materials.  First I will describe two new defect phases,
akin to the Renn-Lubensky twist-grain-boundary (TGB) phase of chiral smectics \TGB ,
the polymer TGB phase and the moir\'e phase \KN .  In the next section I will
describe a three-dimensional hexatic phase with chiral bond order \LCBO.  Finally,
I will discuss recent experiments on DNA \STREY\ which might be interpreted as
true, three-dimensional hexatic phases.

\newsec{Novel Phases of Chiral Liquid Crystalline Polymers}

A notable feature of biological materials is the profusion of long polymer molecules
with a definite handedness.  DNA, polypeptides (such as poly-$\gamma$-benzyl-glutamate) 
and polysacharides (such as xanthan) can all be
synthesized with a preferred chirality.  
Long polymers in dense solution often crystallize into a hexagonal columnar phase.
When the polymers
are chiral this close packing into a triangular lattice 
competes with the tendency for the polymers to twist macroscopically \BOUii\
as in cholesteric liquid crystals.  Similar to the twist grain boundary
phase of chiral smectics \TGB , macroscopic chirality can proliferate when screw 
dislocations enter the crystal. 
Like flux lines in
a type II superconductor, dislocations only appear provided the free energy reduction 
from the chiral couplings
exceeds the dislocation core energy.  If the chirality is weak, a defect
free hexagonal columnar phase persists, as in the Meissner phase of superconductors
\KN.

The hexagonal columnar phase of liquid crystals has broken rotational
invariance around all three co\"ordinate axes.  The associated Goldstone
modes are simply the deviations in the nematic director field away from
the $\bf\hat z$-axis, $\delta\vec n\equiv {\bf\hat n}-{\bf\hat z}$ and
the hexatic \HN\ bond-angle field $\theta_6$ defined in the $xy$-plane \LCBO . 
Fluctuations of the nematic director are controlled by the Frank free energy:
\eqn\ei{\free_{{\delta\vec n}} = {K_1\over 2}\left(\nabla\cdot{\delta\vec n}\right)^2 
+ {K_2\over
2}\left[{\bf\hat z}\cdot(\nabla\!\times\!{\delta\vec n}) - q_0\right]^2 + {K_3\over
2}\left[\partial_z{\delta\vec n}\right]^2}
where $K_i$ are the Frank elastic constants and $2\pi/q_0$ is the equilibrium 
cholesteric
pitch.  The hexatic director is governed by the anisotropic free energy:
\eqn\eii{\free_{\ts} = {K_A^{||}-K_A^\perp\over 2}\left(
\partial_z\ts\right)^2 
+ {K_A^\perp\over 2}\left(\nabla\ts\right)^2 - K_A^{||}\tilde q_0{\bf\hat z}
\cdot\nabla\ts}
where $K_A^{||}$ and $K_A^\perp$ are the spin-stiffnesses parallel and 
perpendicular to the nematic axis, respectively.  Note that the last term is
chiral and is allowed by symmetry: the sign of changes in $\ts$ must be measured
with respect to a vector (the right-hand-rule requires a thumb) which we
choose as the nematic director ${\bf\hat n}\approx {\bf\hat z}$.  Under the
nematic inversion ${\bf\hat n}\rightarrow -{\bf\hat n}$ the sign of $\ts$ will
likewise change and thus ${\bf\hat n}\cdot\nabla\ts$ respects the symmetry
of the phase.  In addition there are are additional non-chiral couplings between
$\delta\vec n$ and $\ts$ \refs{\TON,\GIA}:
\eqn\eiii{\free_{{\delta\vec n}\ts}= 
\bar C\left(\partial_z\ts\right)\left[{\bf\hat z}\cdot(\nabla
\!\times\!{\delta\vec n})\right] +\bar C'\nabla\ts\cdot\nabla\!\times\!{\delta\vec n}}

In addition, the
hexagonal columnar phase is a two-dimensional crystal and thus has
a two-dimensional displacement field $\vec u$ arising from the broken
translational invariance.  Rotational invariance dictates the allowed couplings
and terms in the free energy density \KN:
\eqn\efre{\free_{\rm crystal} = \mu(\partial_iu_j -\epsilon_{ij}\ts)^2 
+ {\lambda\over 2}u_{ii}^2 + \mu'(\partial_z u_i-\delta n_i)^2}
where $u_{ij}\equiv{1\over 2}(\partial_iu_j+\partial_ju_i)$ is the
two-dimensional strain tensor, $\mu$ is the two-dimensional shear modulus, 
$\lambda$ the two-dimensional
bulk modulus, $\mu'$ is the tilt modulus and $\epsilon_{ij}$ is the two-dimensional
anti-symmetric symbol.  The total free energy density is thus
\eqn\efr{\free=\free_{\rm crystal} + \free_{\delta\vec n}
+\free_{\ts}+\free_{\delta\vec n \ts}.}

By minimizing the total free energy with respect to $\delta\vec n$ and $\ts$
we find $\delta\vec n \approx \partial_z\vec u$ and $\ts\approx{1\over
2}\epsilon_{ij}\partial_iu_j$ so that the achiral free energy density is (to
lowest order in derivatives):
\eqn\eee{\free_{\rm eff} = \mu(u_{ij})^2 
+ {\lambda\over 2}u_{ii}^2 + {K_3\over 2}\left(\partial_z^2u_i\right)^2.}
The chiral contribution is subtle:  in terms of the displacement field $\vec u$
the two chiral terms are
\eqn\echi{\free^*_{\rm eff} =
-\gamma\left(\partial_x\partial_zu_y-\partial_y\partial_zu_x\right)
-\gamma'\left(\partial_z\partial_xu_y-\partial_z\partial_yu_x\right),}
where $\gamma=K_2q_0$ and $\gamma'={1\over 2}K_A^{||}\tilde q_0$.
The only difference between the two allowed chiral terms is the order of 
differentiation -- integration by parts on the boundary of the
sample will make these two terms identical!  Moreover, both
terms are total derivatives and thus one may think that they
are unimportant. However, topological defects provide boundaries {\sl inside}
the sample and thus boundary conditions become important in non-uniform
structures.  

When dislocations are introduced 
$\vec u$ is no longer
single valued.  To account for this, one may introduce a new variable
$w_{\gamma i}$ which is equal to $\partial_\gamma u_i$ away from the
defects \KOS .
The free energy becomes
\eqn\efreeii{
F=\int d^3\!x\,{\mu}\left({w_{ij}+w_{ji}\over 2}\right) + {\lambda\over
2}(w_{ii})^2+{K_3\over 2}(\dz w_{zi})^2
- \gamma\epsilon_{ij}\di w_{zj} - \gamma'\dz({1\over
2}\epsilon_{ij}w_{ij}),}
where $\ts={1\over 2}\epsilon_{ij}w_{ij}$ and $\delta n_i=w_{zi}$.

Dislocations are restricted so that the Burger's
vector, $\vec b$ must lie
in the $xy$ plane. 
We introduce the density tensor $\alpha_{\gamma i}({\bf r}) = \int d{\bf t}
d{\vec b}\,
t_\gamma b_i \rho({\bf t},\vec b, {\bf r})$, where $\rho({\bf t},\vec b,{\bf
r})$ is the volume density
of dislocations at the point ${\bf r}$ with Burger's vector $\vec b$ pointing
in
the
$\bf t$ direction. Since the dislocations do not end, $\nabb\cdot {\bf t}=0$, and
$\partial_\gamma\alpha_{\gamma i}\equiv 0$.  Following \KOS\ it is straightforward
to show that:
\eqn\eburg{\epsilon_{\mu\nu\gamma}\partial_\nu w_{\gamma i} = 
- \alpha_{\mu i}.}  This relation can be manipulated to show
that \refs{\IND,\KN}:
\eqn\ior{2\dz\ts-{\bf\hat z}\cdot\nabla\cross\delta\vec n = -\Tr[\alpha].}
This relation shows that screw dislocations, which only contribute to the 
symmetric part of $\alpha$ require either a twisting of 
the director $\delta\vec n$
or the hexatic director $\ts$.  

Using the effective free energy \eee\ one can calculate the energetic penalty for 
screw and edge dislocations in the crystal.  The free energy per unit length
of a screw dislocation is finite (independent of the system size).  Thus
if the desire to twist is large enough defects will proliferate to allow
the system to twist.  We have proposed two new defect phases: the polymer
tilt-grain-boundary phase in which the director twists discretely
across grain-boundaries (this phase is nearly identical in morphology
to the Renn-Lubensky twist-grain-boundary phase \TGB) and the moir\'e phase
in which the bond-order rotates along the polymer axis.  We show a portion
of this structure in Figure 1 and the phase diagram for our system in Figure 2.
Recently, the morphology of the moir\'e phase
has been proposed as the steady-state structure of flux-lines in 
superconductors in a current parallel to the magnetic field \FLUX .
  
\newsec{The Polymer Hexatic}
When the crystalline order melts, the Goldstone modes of broken
rotational invariance, $\delta\vec n$ and $\ts$ become the interesting
degrees of freedom.  The N+6 \TON\ phase of liquid crystals is
thus very similar to a biaxial nematic phase, although instead of
having two-fold symmetry perpendicular to the nematic director, the N+6 phase
has {\sl six-fold} symmetry.  Recently, in a DNA system, 
the NIH group \STREY\ has 
seen the first evidence of a nematic phase with hexatic order.  
I will argue that this is rather surprising:  unless the tendency
to twist around the nematic axis ($\tilde q_0$) is small, 
Landau theory predicts that {\sl either} the nematic order {\sl or} the bond-orientational
order must twist.  If that were the case, the X-ray scattering in the plane
perpendicular to the nematic director would be a powder average over
many different, rotated hexatic regions.  Thus one might expect that there should
be a ring in the $q_{\tiny\perp}$ plane, rather than the observed $\cos 6\theta$ modulation
as shown in Figure 3.

We can recast the hexatic free energy in terms of the hexatic order
parameter $\psi_6$.  When there is no hexatic order, $\psi_6=0$, and when
there is hexatic order $\psi_6=\vert\psi_6\vert\exp{6i\ts}$.  For simplicity
I take $K_A^\perp=K_A^{||}=K_A$.  The free energy
density can be recast as \LCBO :
\eqn\ehexfree{\eqalign{\free &=
 {1\over 2}K_1\left(\nabb\cdot{\bf\hat n}\right)^2
+{1\over 2}K_2\left[{\bf\hat n}\cdot\nabb\cross{\bf\hat n} - q_0\right]^2
+{1\over 2}K_3\left[{\bf\hat n}\cross\left(\nabb\cross{\bf\hat n}\right)\right]^2\cr
&\qquad +\left\vert\left(\partial-i\tilde q_0{\bf\hat
n}\right)\psi_6\right\vert + r\left\vert\psi_6\right\vert^2 +
u\left\vert\psi_6\right\vert^4,\cr}}
where $r$ is the reduced temperature and $u$ is an interaction parameter.  Note
that this theory is {\sl identical} to that for a smectic-A liquid crystal composed
of chiral molecules, first proposed by de Gennes \DG .  The phenomenology
of this model can be borrowed from the theory of superconductors: there are
two possible uniform phases of this system.  There
is a cholesteric phase (normal metal) in which $\psi_6$ 
vanishes and ${\bf\hat n}\cdot\nabb\cross{\bf\hat n} = q_0$ 
and the chiral hexatic phase (Meissner phase) 
in which $\psi_6$ is
non-zero and the nematic director points along a single axis.  Note that in
this phase the hexatic director must rotate about the nematic axis with 
an inverse pitch $2\pi/\tilde q_0$, as shown in Figure 4.  

Finally we note that if the system has a stable defect phase ({\sl i.e.}, it is
like a type-II superconductor) then it can have a Renn-Lubensky TGB phase
in which grain-boundaries separate regions of perfect chiral hexatic order.

\newsec{Where Is The Twist?}
Unfortunately, the above analysis suggests that when chiral molecules
form N+6 phases, either the hexatic order must twist or the nematic director
must twist.  The data presented in Figure 3 is contrary to this result -- if
the hexatic order were twisting, the 6-fold modulation would be washed out
and turned into a ring.  Since this same DNA system, at lower concentrations,
forms a cholesteric with a micron-sized pitch, it is hard to imagine that 
the hexatic order does not also twist at the micron scale.  The illuminated
area in the X-ray experiment is on the order of one millimeter, so one
would expect many twists of the hexatic order.  

One possibility is that the DNA stiffness makes twisting difficult \ILCC .  
Unfortunately, since the persistence length is $50$ nanometers, 
DNA molecules can easily bend around each other on 
the micron length scale.  Another
possibility is that fluctuations can reduce the chiral strength.  This
is known to happen in polymer cholesterics \refs{\TLL,\KT} .  However, even
with the most optimistic estimates, this
effect produces only a factor of two increase of the pitch.  Finally, one might
calculate the values of $q_0$ and $\tilde q_0$ via a microscopic approach \HKL .
This work is in progress.

\newsec{Acknowledgements}
The work on columnar phases was done in collaboration with 
D.R.~Nelson.  Conversations with him, P.~Aspinwall, A.B.~Harris, 
T.C.~Lubensky, H.~Strey, V.A.~Parsegian, R.~Podgornik and J.~Toner
are gratefully acknowledged.  
The author is supported by NSF CAREER Grant DMR97-32963, an award from
Research Corporation and a gift from L.J. Bernstein.

\listrefs
\listfigs

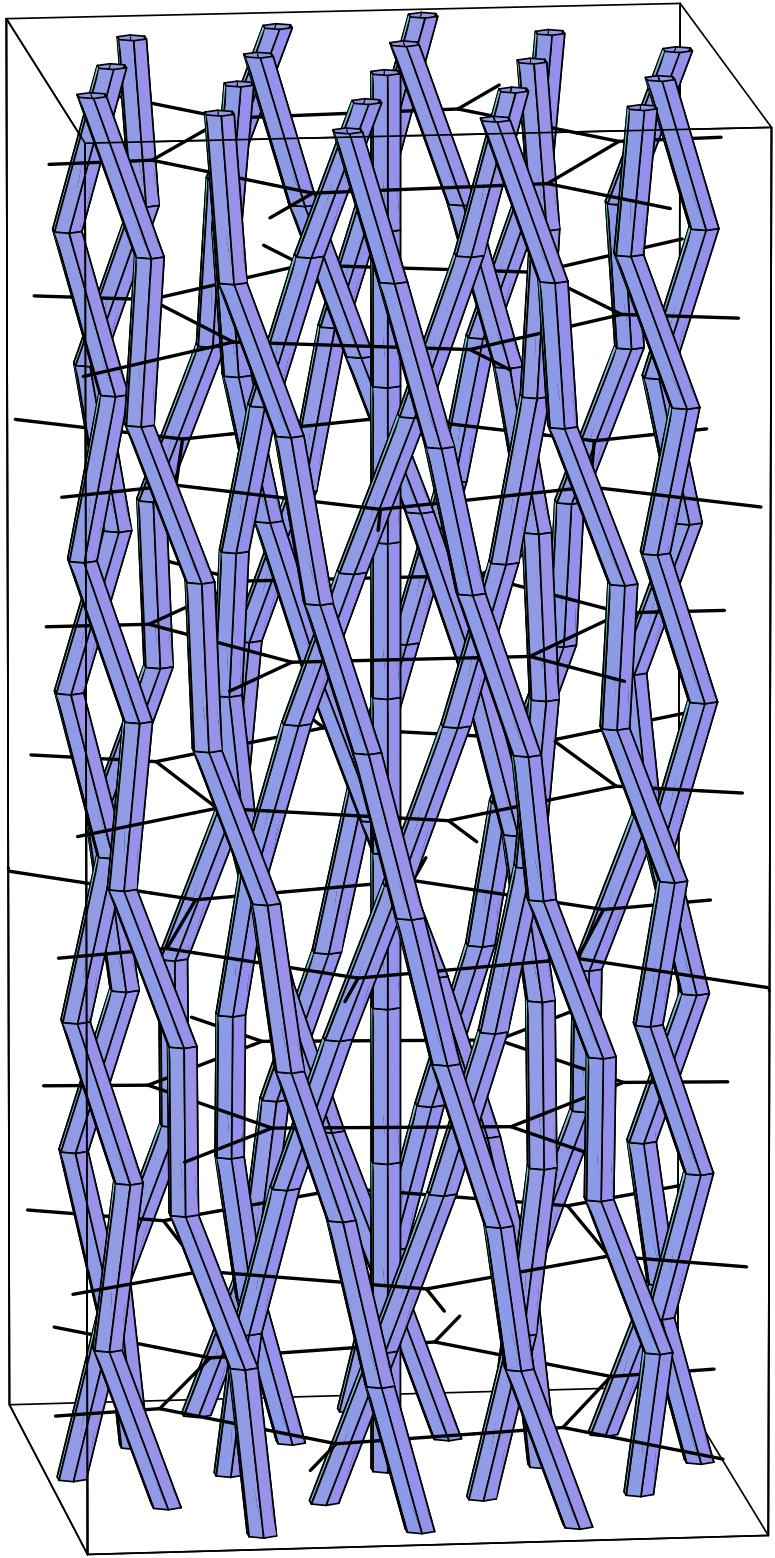
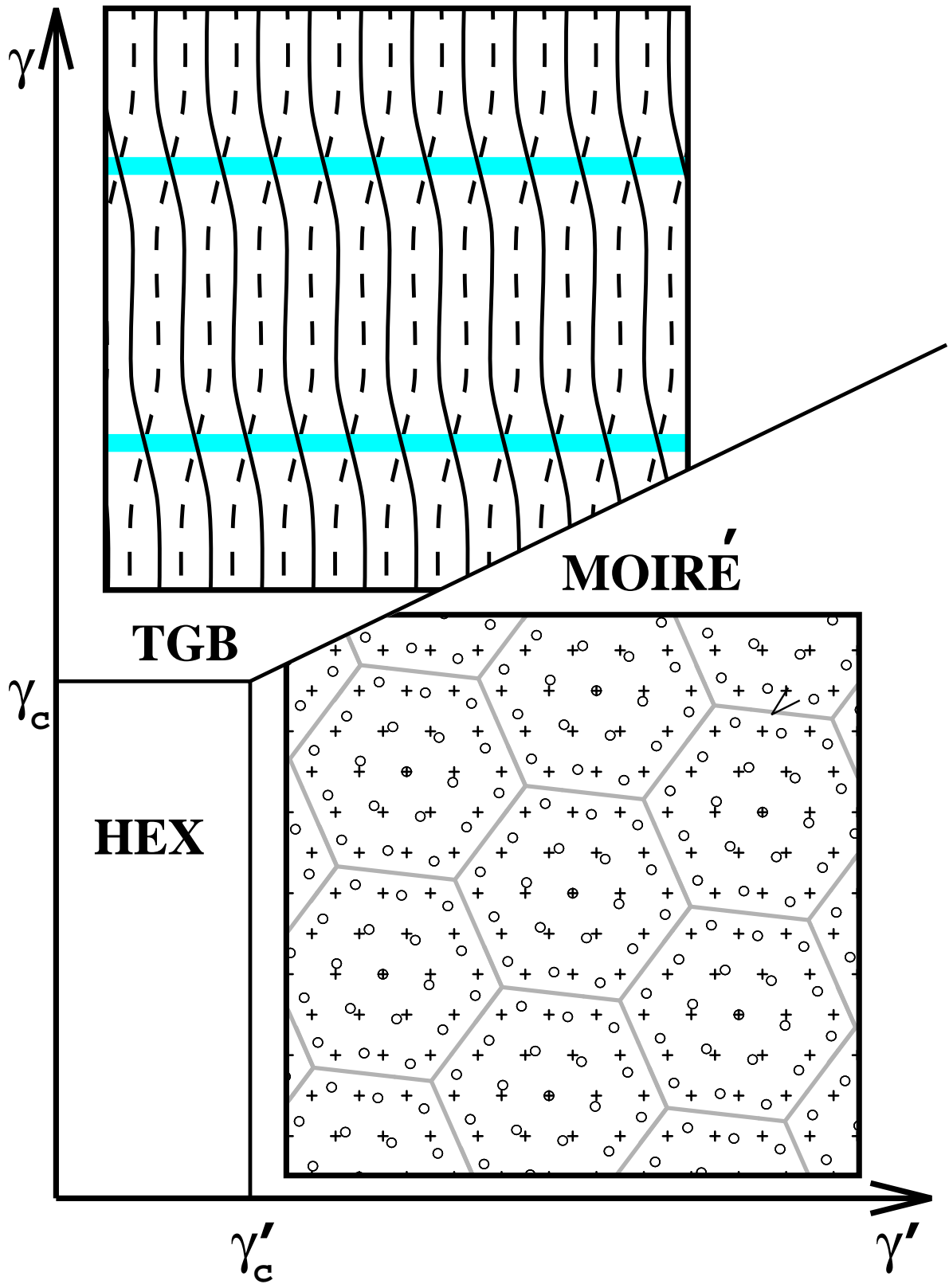
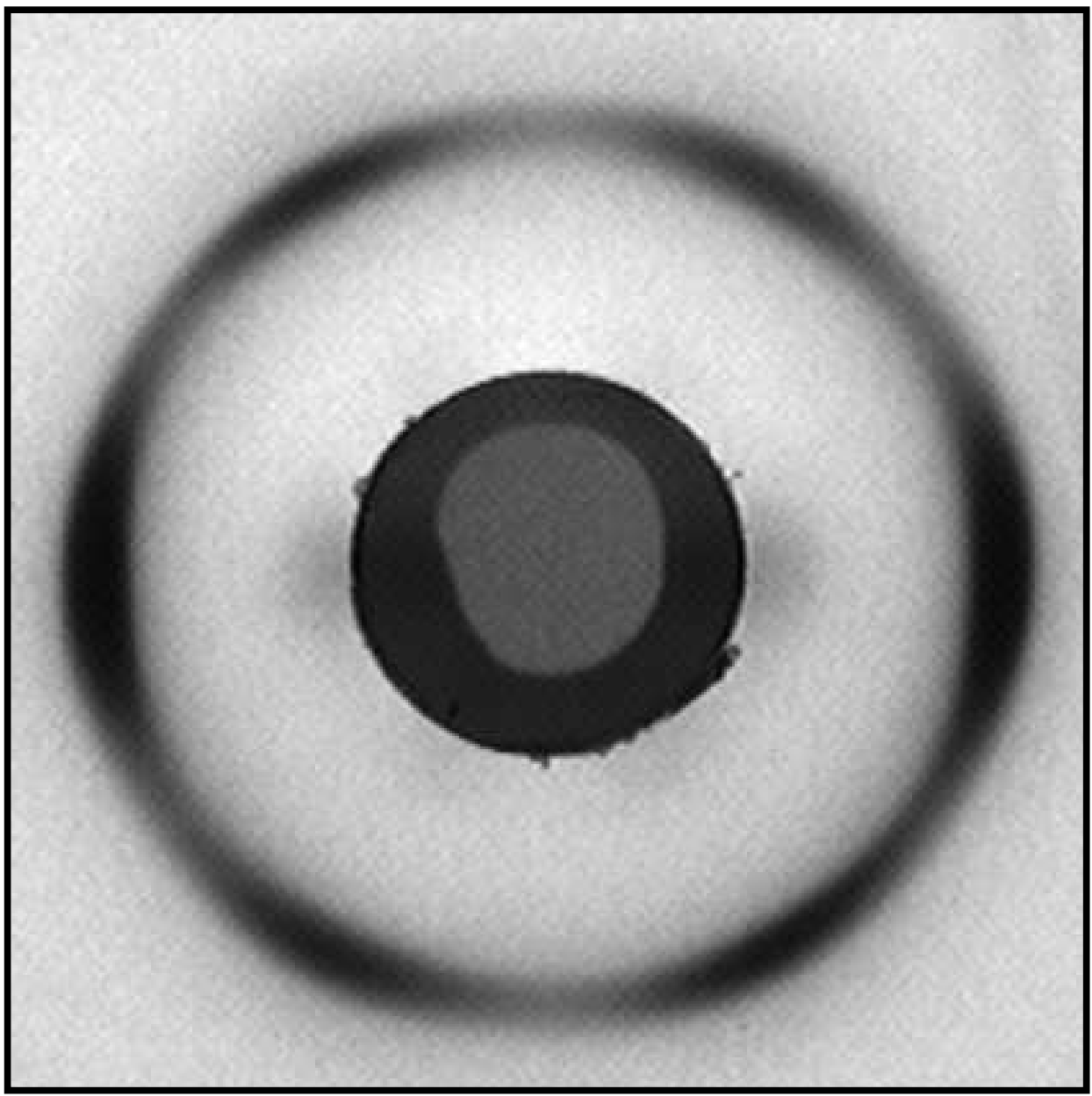
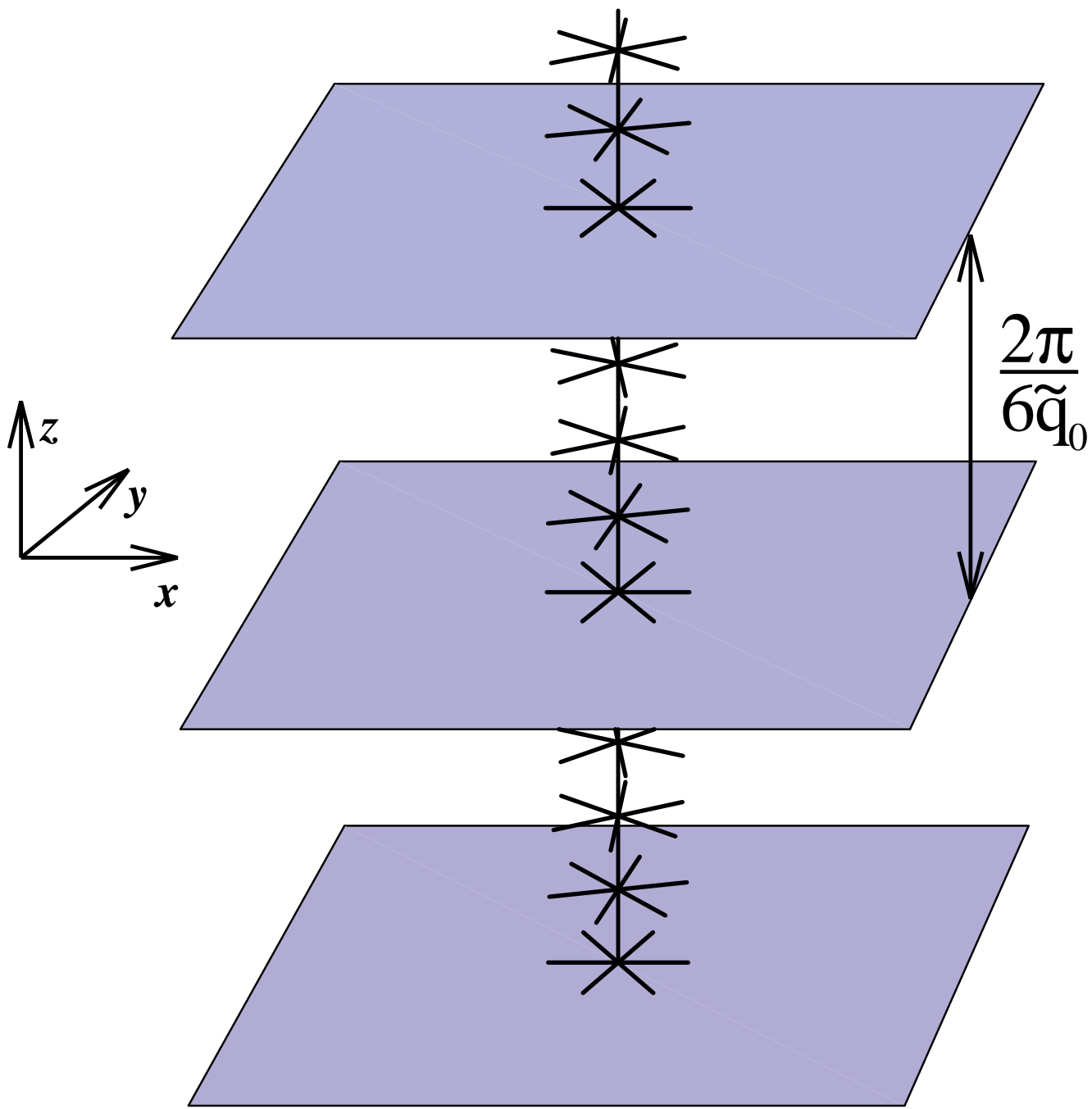
\bye